\documentclass[10pt,a4paper]{article}
\usepackage{graphicx,color}
\begin{document}

\title{Optimized focusing ion optics for an ultracold deterministic single ion source targeting nm resolution}

\author{Robert Fickler$^1$, Wolfgang
Schnitzler$^1$, Norbert M. Linke$^2$,\\ Ferdinand Schmidt-Kaler$^1$, Kilian Singer$^1$$^{\ast}
$\thanks{$^\ast$Corresponding author. Email:
kilian.singer@uni-ulm.de\vspace{6pt}} \\ \\
\vspace{6pt}{\em{$^1$Institut f\"ur Quanteninformationsverarbeitung,
Universit\"at Ulm, Ulm, Germany}}\\ 
\vspace{6pt}{\em{$^2$Department of Physics, University of Oxford, Clarendon Laboratory,  Oxford, U.K.}}\\} \vspace{6pt}

\maketitle

\begin{abstract}
Using a segmented ion trap with~mK laser-cooled ions we have
realised a novel single ion source which can deterministically
deliver a wide range  of ion species, isotopes or ionic molecules
[Schnitzler et al., Phys. Rev. Lett. 102, 070501 (2009)].
Experimental data is discussed in detail and compared with numerical
simulations of ion trajectories. For the novel ion source we
investigate numerically the influence of various extraction
parameters on fluctuations in velocity and position of the beam. We
present specialized ion optics and show from numerical simulations
that nm resolution is achievable. The Paul trap, which is used as a
single ion source, together with the presented ion optics,
constitutes a promising candidate for a deterministic ion
implantation method for applications in solid state quantum
computing or classical nano-electronic devices.

\begin{center}
\end{center}
\end{abstract}

\section{Motivation}
Over the last few years, integrated semiconductor devices have
reached structure sizes in the order of a few tens of nanometres,
and further miniaturization is expected. Thus it is becoming more
and more important to dope the devices in an exact predetermined and
reproducible manner. In the next few years, the amount of doping
atoms in the active region of a field effect transistor might drop
below 100, then
 statistical Poissonian fluctuations which arise from
conventional doping techniques will be significant. At nanometre
length scales, only small fluctuations in the number of doped atoms
are sufficient such that the assumption of homogeneously distributed
doping atoms is no longer valid and the electronic characteristics
are disturbed \cite{SHINADA2005}. But not only conventional solid
state devices would benefit from an accurate quantity of doping
atoms, a future solid state quantum computer fully relies on
precisely placed single dopant atoms; at well defined separation
distances and depths. In those quantum devices, single embedded
impurity atoms, e.g. phosphorus dopants in silicon \cite{KANE1998}
or colour centres in diamond \cite{GREENTREE2008}, are used to
retain and process information in a quantum mechanical way. Nowadays
it is possible to address single quantum devices like
nitrogen-vacancies (NV) colour centres in diamond and manipulate
them coherently over several~$\mu$s or even ms
\cite{GURUDEV2007,NEUMANN2008}. However, when it comes to scalable
quantum computers with more than a few qubit carriers, one of the
most important challenges is to place the dopant atoms at an exact
position and with uniform separation at nm resolution. One method
proposed for fabrication of these assemblies uses lithography based
on scanning tunneling microscopy (STM)
\cite{OBRIEN2001,SCHOFIELD2003,RUESS2004,POK2007,RUESS2007}. A
hydrogen terminated silicon surface is structured with an STM,
followed by chemical reactive surface binding of the doping atoms.
Although the positioning of the incorporated single phosphorous
dopants is realised with sub-nm accuracy, the technique is limited
to silicon surfaces and unavoidable impurities in the background gas
can lead to functional impairment. Another method for controlling
the amount of doping atoms is the direct implantation of atoms or
ions. Here, the common approach utilizes statistical thermal sources
which provide a dense ion beam that has to be thinned out by several
choppers and apertures. To ensure single ion implantation it is
necessary to detect the implantation event by observing the
generated Auger electrons, photoluminescence, phonons, electron-hole
pairs or changes in the conductance of a field effect transistor
\cite{SHINADA2002,PERSAUD2004,MITIC2005,BATRA2007,SHINADA2008}.
Therefore, the implantation only works if either the ions are highly
charged or if they are implanted with large kinetic energies. With
both systems it is possible to achieve a resolution of less than
10~nm, but their use is limited to cases where either highly charged
states or high kinetic energies are available, and therefore deep
implantation is unavoidable. Both methods lead to surface damages
and additional inaccuracies in depth and lateral position due to
statistical straggling. Our method is universally applicable to a
wide range of doping atoms and it allows implanting at very low
energies, thus avoiding the problems described above.

\section{Ultracold deterministic ion point source}
We have realised a novel system for direct implantation of ions into
solid state surfaces by using a Paul trap as an ultracold
deterministic ion source
\cite{MEIJER2006,MEIJER2008,SCHNITZLER2009}.

\emph{Specialized linear Paul trap:} Central component of our
technique is a linear segmented Paul trap with laser cooled
$^{40}\mathrm{Ca}^+$ ions similar to those utilised for scalable
quantum information processing \cite{ROWE2002}. A Paul trap is a
well known tool for trapping single, charged particles by using
static (dc) electric fields and an alternating (rf) field thereby
producing a pseudo-potential of around 1 eV depth. With various
laser cooling techniques, the trapped ions can be cooled to the
motional ground state \cite{ROOS2000,KING1998}. In addition, it is
possible to trap other charged particles or even molecules that
cannot be directly laser cooled but can be sympathetically cooled
due to their electrostatic interaction. Identification of those
additionally loaded doping ions, which are invisible to laser light,
can be conducted by exciting collective vibrational modes
\cite{NAEGERL1998,DREWSEN2004}. The segmented trap design is capable
of separating and transporting ions over a distance of a few cm
\cite{HUBER2008}. Our design consists of four copper plated
polyimide blades of 410~$\mu$m thickness and 65~mm length arranged
in an x-shaped manner with a distance of 2~mm between opposing
blades \cite{SCHNITZLER2009}. The dc voltages are applied to eight
segments of 0.7~mm width on the top and bottom of each blade. A
unique feature of our design is that it is capable to shoot out ions
in a well defined axial direction by switching two of the electrode
segments to a higher voltage. Unlike conventional linear Paul traps,
our design does not lose its radial confinement even when biasing
the dc trap potential to high voltages. This is realised by applying
the rf to the electrodes at the inner front edges of two opposing
blades, while the other two are grounded. Furthermore, the trap
provides an additional broader deflection electrode on every blade,
which is used to aim the extracted ions in the demanded direction.
This special arrangement enables the exact axial extraction of the
cooled $^{40}\mathrm{Ca}^+$ ions and sympathetically cooled dark
ions due to the preservation of the radial confinement during the
extraction sequence. The development of this special design was only
feasible with our custom designed simulation software, see
sect.~\ref{sim}. Characteristic working conditions are an rf voltage
with an amplitude of 200~V at a frequency of $\Omega/2\pi$ =
12.155~MHz, which yields a radial secular frequency
$\omega_\mathrm{rad}/2\pi$ = 430~kHz for a $^{40}\mathrm{Ca}^+$ ion.
The required dc potential is generated by a voltage of 35~V which is
applied to trap segments 2 and 8 and leads to a frequency of the
axial potential of $\omega_\mathrm{ax}/2\pi$ = 280~kHz. The vacuum
chamber itself is made out of stainless steel and evacuated down to
a base pressure of $3\times10^{-9}$~mbar by a turbo molecular pump,
an ion-getter pump and a titanium sublimation pump. Calcium and
dopant ion generation is induced via a multiphoton process e.g. by a
pulsed frequency tripled Nd-YAG laser with a wavelength of 355~nm
and a pulse power of 7~mJ.

\emph{Observation of ions:} With this setup the trapped
$^{40}\mathrm{Ca}^+$ ions are located above segment 5, where they
are Doppler cooled and illuminated by laser light near 397~nm,
866~nm and 854~nm. In order to image single ions or ion-crystals we
collect the scattered photons by a f/1.76 lens on an electron
multiplying charge coupled device (EMCCD) camera. While the calcium
ions can be directly cooled by the installed laser system and
identified on the image from the camera, the additionally loaded
doping ions are sympathetically cooled and can be easily identified
by the voids in the image of the ion crystal \cite{SCHNITZLER2009}
as they are invisible to the irradiated laser light. By applying an
ac voltage to an electrode located under the ion crystal we can
stimulate collective vibrational modes with characteristic resonance
frequencies $\omega_\mathrm{ax}$ and therefore specify the exact
species of these dark ions \cite{NAEGERL1998}. Another determination
of the additionally loaded doping ions can be implemented by
measuring the mass ratio with amplitude modulated resonant laser
light \cite{DREWSEN2004}. So far, the trapped particles are simply
Doppler cooled down to $T=2$~mK, which is slightly above the Doppler
limit and was deduced from the width of the excitation spectrum on
the $S_\mathrm{1/2}$ - $P_\mathrm{1/2}$ laser cooling transition.
With further cooling methods using electromagnetically induced
transparency, which is currently implemented in our lab, it is
possible to cool the broad band of vibrational frequencies of an ion
crystal simultaneously and thereby reach the motional ground state
\cite{ROOS2000,KALER2001}. Under these initial conditions the ion
trap would operate at the fundamental limit given by the Heisenberg
uncertainty relation. In order to exclusively implant doping ions
they can be separated from the cooling $^{40}\mathrm{Ca}^+$ ions by
either splitting the ions inside the trap before the extraction is
performed or after extraction by deflecting unwanted ions e.g. by
increasing the voltage of an einzel-lens. The separation of the
$^{40}\mathrm{Ca}^+$ ions and the doping ion inside the trap can be
achieved by converting the axial potential into a double well and
transporting the doping ion away by subsequently applying time
dependent dc voltages \cite{HUBER2008}.

\emph{Extraction of ions:} The extraction process itself is induced
by biasing segments 4 and 5 with 500~V (supplied by iseg inc., Model
EHQ-8010p) within a few nanoseconds. This fast switching is achieved
by two high voltage switches (Behlke inc., HTS 41-06-GSM) which are
triggered by a TTL signal from our control computer. In addition, we
synchronized the switching event with the rf field phase in order to
reduce fluctuations of velocity and position of the ion by
implementing an electronic phase synchronization circuit which
delays the TTL signal to a well defined constant time period after
the zero crossing of the rf. The experimentally measured standard
deviation of the delay time equals 0.34~ns. For the detection of the
extracted ions or dopants we use an electron multiplier tube (EMT)
with 20 dynodes (ETP inc., Model AF553) which is supplied with a
voltage of -2.5 kV and has a specified quantum efficiency of about
80\% for positively charged particles. The gain is specified with 5
$\times$ 10$^5$ and provides a 100 mV electrical signal which is 10
to 15~ns wide for each detection event. The EMT is accommodated in a
second vacuum chamber which is located at 247~mm from the trap and
can be completely separated by a valve to facilitate prospective
changes of target probes.

\section{Numerical simulation of the ion source and specialized ion optics}\label{sim}
For designing an optimised trap and ion optics we developed a
simulation software package which is based on a fast multipole
solver \cite{GREENGARD1988,NABORS1994} with additional high accuracy
solvers respecting symmetry properties of the trap and the ion
optics. The software allows deducing accurate electrostatic
potentials from any CAD-model. With this simulation software we
describe the trajectories during the extraction and the expected
spots at the target. We analyse the dependence of the spatial
dimension and the velocity fluctuation of the single ion beam on the
initial ion temperature, the start position inside the trap and the
trigger phase of the rf voltage at the extraction event.
Additionally, we compare different designs of possible einzel-lenses
and study the simulated focusing properties as well as the
possibility of our system to correct spherical abberation of the
lens and to filter the cooling ions during the implantation process.
To verify our simulation with experimental data we have checked
different trap geometries and found an agreement of the axial and
radial trapping frequencies within 2 to 3 \%.

\begin{figure}[htb]
\begin{center}
\includegraphics[width=13cm]{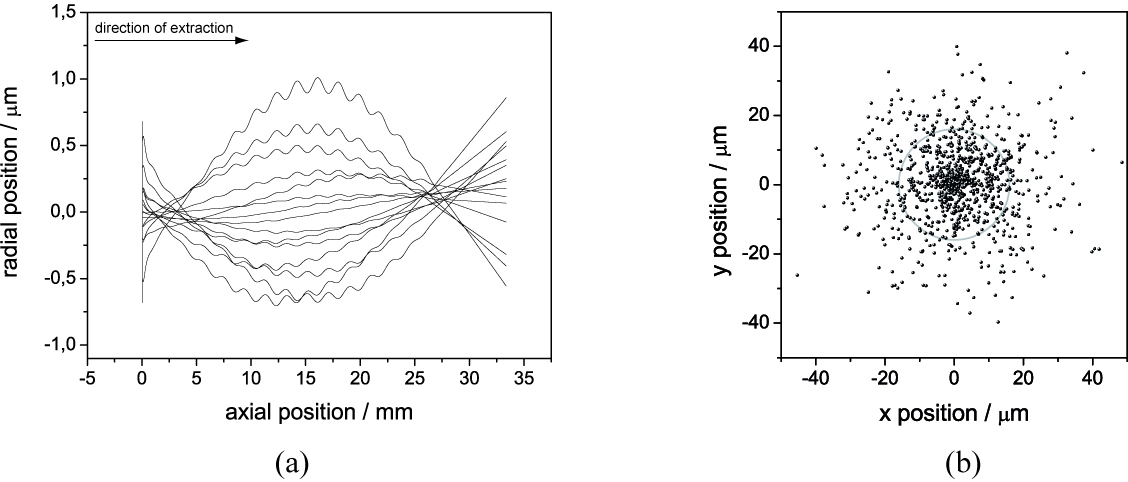}
\caption{\setlength{\baselineskip}{11pt}{\small(a) Trajectories of $^{40}\mathrm{Ca}^+$ ions during
extraction from the trap with an initial temperature of 2~mK. (b)
Resulting spot diagram at a distance of 247~mm between trap centre
and target amounts to a beam divergence of 134~$\mu$rad (light grey
circle illustrates the 1$\sigma$-spot radius of 16~$\mu$m).}}
\label{fig:fig1}
\end{center}
\end{figure}
\subsection{Deterministic ion point source}
For an exact simulation of the trap potentials we have sketched the
complete trap geometry as a full three dimensional CAD-model. By
virtually applying the same voltages as we do in the real experiment
(values mentioned above) we are able to calculate the trajectories
of the trapped particles with full time dependent dynamics including
micro motion at frequency $\Omega$. The initial momentum and
position of the ion inside the trap is determined from the thermal
Boltzmann distribution at a selectable temperature. The extraction
process is triggered by switching two segments on every trap blade
to 500~V synchronized to a well defined phase of the rf voltage. The
plotted trajectories show the influence of the micro motion during
the extraction and the resulting spread of the ion beam (see
Fig.~\ref{fig:fig1}(a)). Slight numerical asymmetries, which appear
despite the symmetrization procedure, are corrected by small
compensation voltages on the rails as in the real experiment.

\emph{Dependence of the ion beam on the initial ion temperature:}
With an initial ion temperature of 2~mK the results show a
1$\sigma$-spot\footnote{Note that the 1$\sigma$-expression is always
used to express that 68\% of the studied data lies within the given
interval, although the some results are not perfectly reproduced by
a Gaussian distribution.} radius of 16.5~$\mu$m at a distance of
247~mm between trap centre and target which amounts to a full angle
beam divergence of 134~$\mu$rad (see Fig.~\ref{fig:fig1}(b)). The
specified temperature of 2~mK is similar to the experimentally
achieved value and the deflection electrodes are grounded during the
extraction simulation. The mean longitudinal velocity of the
extracted ions at the target is calculated to be 22.1~km/s with an
1$\sigma$-uncertainty of only 1.3~m/s. Therefore, the predicted ion
beam shows promising characteristics for subsequent focusing with
ion optics due to small spherical and chromatical aberration. The
predictions are even better if sub-Doppler-cooling methods are
applied. This would lead to mean phonon expectation values below 1
and temperatures of 100~$\mu$K for the ions inside the trap can be
achieved. Then the setup would work at the Heisenberg limit. The
full angle divergence of the generated ion beam shrinks to
30~$\mu$rad. This means the 1$\sigma$-spot size at the target
amounts to 3.7~$\mu$m and therefore the beam improves significantly.
However, as the extraction voltage remains the same, the mean
longitudinal velocity is not influenced by the smaller initial
temperature, but the 1$\sigma$-velocity uncertainty decreases
slightly to 1~m/s. Our simulations show that by applying voltages to
the deflection electrodes, the characteristics of the resulting spot
shapes change enormously. For example, if 7.5~V between two opposing
deflection electrodes and 9.1~V between the others are applied
respectively (which resembles our experimental setup), the spot is
approximately shifted 19.4~mm in the horizontal and 30.5~mm in the
vertical direction. In addition to this expected deflection the spot
is stretched from the former Gaussian to a cigar-shaped distribution
and therefore the 1$\sigma$-spot size value cannot be calculated
(see Fig.~\ref{fig:fig3}(b) zoomed area). This drastic change is
simulated for 2~mK cooled ions as well as for sub-Doppler-cooled
ions.

\begin{figure}[htb]
\begin{center}
\includegraphics[width=14cm]{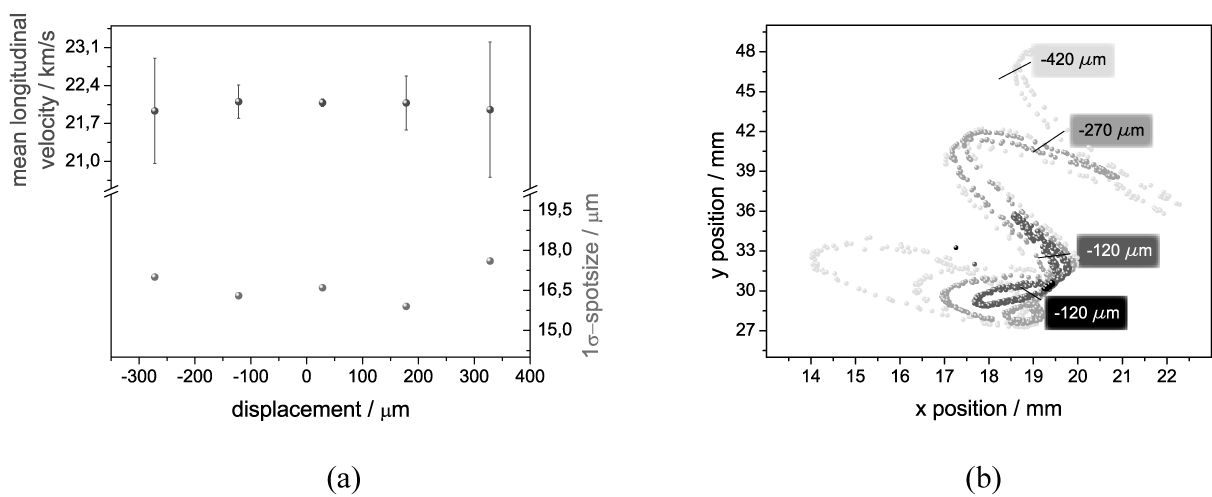}
\caption{\setlength{\baselineskip}{11pt}{\small Dependence of the ion beam on the start position (positive
values mean towards the extraction direction). (a) Resulting mean
longitudinal velocities and 1$\sigma$-spot sizes after extraction at
a distance of 247~mm from the trap centre with deflection electrodes
grounded. (b) Resulting spot diagrams of extracted ions when the
deflection electrodes are biased to 7.5~V between two opposing trap
blades and 9.1~V between the other two blades respectively.}}
\label{fig:fig2}
\end{center}
\end{figure}
\emph{Dependence of the ion beam on the initial start position:}
Aside from the dependence on the initial temperature, the
transversal expansion and the velocity of the ion beam with its
uncertainty strongly depends on the start position of the extracted
ion. Whereas altering the initial position does not change the size
and the (Gaussian) spot shape at the target significantly when
deflection electrodes are grounded, the velocity fluctuation is
strongly affected (see Fig.~\ref{fig:fig2}(a)). For displacements up
to $\pm$400~$\mu$m the mean longitudinal velocity for the different
displacements has a constant value of 22~km/s ($\pm$ 2 \%). On the
contrary, the velocity spread at each displacement varies
enormously: from only 1.3~m/s when the ions are extracted from the
theoretical minimum of the axial potential, it increases up to
1.5~km/s when they are shifted around 420~$\mu$m out of the centre
of the potential before being extracted. This is caused by a strong
increase of the micro motion of a trapped ion when it is no longer
in the minimum of the trapping potential, and therefore has a highly
varying initial momentum. For the 1$\sigma$-spot sizes it is also
advantageous to extract the ion out of the potential minimum. Then,
a 1$\sigma$-spot of about 16 $\mu$m can be achieved. However, a
displacement of $\pm$400~$\mu$m only leads to an enlargement of the
1$\sigma$-spot size by additional 2~$\mu$m. The influence of the
start position on the ion beam properties becomes more crucial when
the deflection electrodes are supplied with a non zero voltage. Due
to slightly displaced trajectories during the extraction process,
the ions begin to oscillate by reacting to the rf field and
therefore lose their well-defined extraction direction. In Figure
\ref{fig:fig2}(a) the ions are displaced up to 420~$\mu$m from the
potential minimum before extraction which means that the ion is
trapped in the middle of two electrodes. The former cigar shaped
extension of the spot of around half a millimetre increases
tremendously to about 10 to 20~mm. Additionally, the shape changes
to a screw like structure (see Fig.~\ref{fig:fig2}(a)) and
deterministic aiming becomes nearly impossible. The mean value of
the longitudinal velocity sums to around 22~km/s and changes only by
2 \% for different extraction positions, while fluctuations increase
to a maximum of 1.6~km/s for ions shifted 420~$\mu$m out of the
trapping potential minimum.

\begin{figure}[htb]
\begin{center}
\includegraphics[width=14cm]{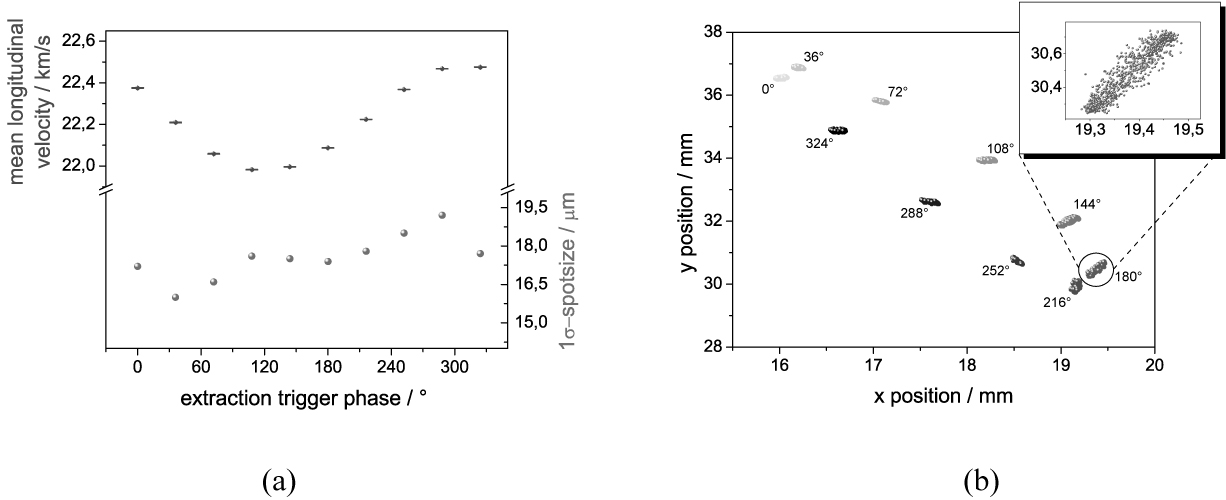}
\caption{\setlength{\baselineskip}{11pt}{\small Dependence of the ion beam on the phase of the rf voltage.
Initial temperature of the ions inside the trap is kept constant at
2~mK and the distance is set to 247~mm from the trap centre for all
shown results. (a) Resulting mean longitudinal velocities and
1$\sigma$-spot sizes with deflection electrodes grounded. (b)
Resulting spot diagrams of extracted ions with deflection electrodes
are biased to 7.5~V between two opposing trap blades and 9.1~V
between the other two blades respectively. In (b) one spot is zoomed
in to show that a Gaussian distribution is no longer an accurate
assumption if the ions are deflected during extraction.}}
\label{fig:fig3}
\end{center}
\end{figure}
\emph{Dependence of the ion beam on the phase of the rf voltage:}
Similar results are calculated for the augmentation of the beam
expansion while changing the phase of the rf voltage at the time of
the extraction event (see Fig.~\ref{fig:fig3}(b)).
Here, the 1$\sigma$-spot size varies from 16~$\mu$m to 19~$\mu$m
over a whole period of the radio frequency voltage but remains
Gaussian distributed. The mean longitudinal velocity oscillates in a
sine-like manner over an interval of 0.5~km/s around 22.2~km/s
during the different trigger phases. The velocity spread only varies
negligibly between 1.3~m/s and 2.3~m/s. This shows that with
deflection electrodes grounded, the trigger phase at the extraction
event is important to reach a single ion beam featuring the
promising characteristics of a narrow velocity fluctuation and a
small beam divergence. Again, even more important becomes the exact
triggering phase of the extraction event when the deflection
electrodes are used to aim the ions during the extraction process.
In Figure \ref{fig:fig3}(b) the simulated spots are shown for
altering the trigger phase and deflecting the ions with the same
voltages mentioned above. The spot shape for a constant trigger
phase maintains its oval shape and size. When changing the phase the
spots are distributed over an area of 10~mm by 15~mm. The mean
velocity for each different trigger phase stays within 1 \% at
around 22~km/s, and the fluctuations only add up to around 2.5~m/s.
Hence, to achieve a beam of single deterministic extracted ions
suitable for implantation it is important to align the trap axis to
the ion optic axis as accurately as possible and therefore avoid the
usage of the deflection electrodes. Besides the alignment, it is
also crucial to position the ion into the radial pseudo-potential
minimum. Furthermore, the extraction itself has to be synchronised
to the rf field as exactly as possible.

\subsection{Ion optics}
In order to use our specially designed ion trap as a deterministic
implantation tool it does not suffice to shoot the ions out of the
trap with the aforementioned small values of beam divergence and
velocity fluctuations: the ion trajectories have to be focused down
to a few nm. To realise this goal we have developed ion optics which
should be able to focus the beam down to the required nanometre
regime, even to a 1$\sigma$-spot size of around one nm without any
aberration corrections. Due to the narrow ion trajectories and the
low fluctuations in the longitudinal velocity of the ions, the
focusing optics can be kept simple without sophisticated aberration
corrections
\cite{WEISSBAECKER2001,WEISSBAECKER2002,SZEP1988,HAWKES1989}. A
rotationally symmetric simple electrostatic einzel-lens yields
adequate results. This type of ion lens consists of three electrodes
where the first and the third electrode are on the same potential.
The electrostatic potential resembles a saddle surface and can be
generated in two different modes, decel-accel and accel-decel mode.
The former mode is achieved when the middle electrode is biased to a
voltage with the same sign as the charged particle (in our case a
positive voltage) while the lens operates in the accel-decel mode
with a voltage of the opposite sign of the charged particles. Both
modes offer different advantages \cite{LIEBL2008}. In the
accel-decel mode the undesirable chromatical and spherical
aberrations are smaller than in the decel-accel mode. The chromatic
aberration is reduced due to higher velocities of the particles
inside the lens and a therefore lower relative velocity spread
$\Delta$v/v. Additionally, spherical aberration is smaller because
the ion trajectories are closer to the optical axis during their
passage through the lens \cite{SHIMIZU1983} since they are always
deflected towards the axis,while in the decel-accel mode the ions
diverge at first before being focused to the centre. On the other
hand, the decel-accel mode offers three essential advantages.
Firstly, the lens requires lower centre electrode voltages for
reaching a similar focal length. In addition, by applying positive
voltages it is possible to correct spherical aberration by switching
the voltage to a higher potential while the particle is going
through the lens \cite{SCHOENHENSE2002} and in our setup the lens
can be used as a deflector for different ion species (both will be
discussed more in detail below).

\begin{figure}[htb]
\begin{center}
\includegraphics[width=15.5cm]{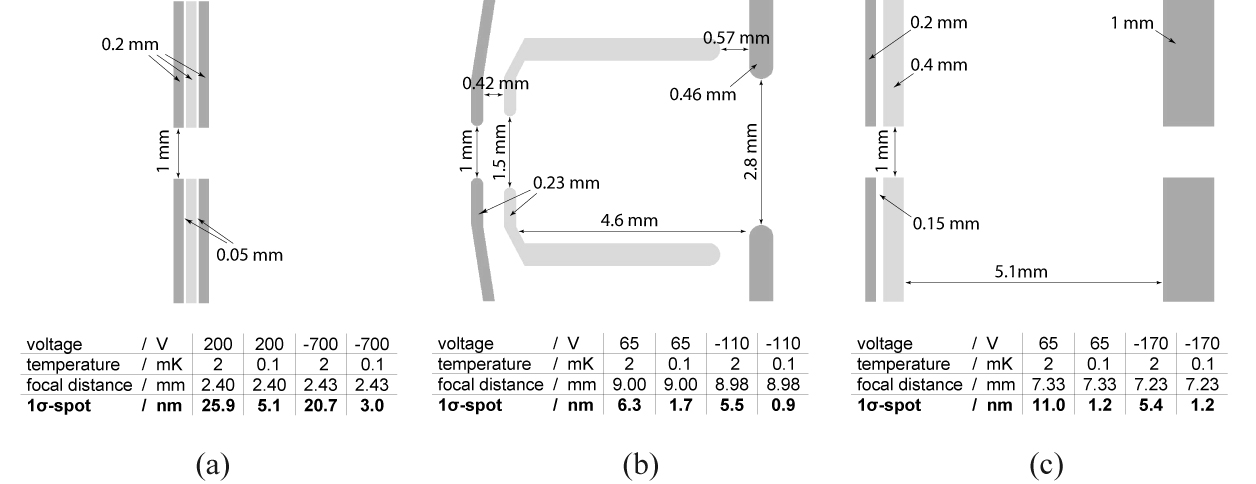}
\caption{\setlength{\baselineskip}{11pt}{\small Comparison of three different shapes of einzel-lenses.
Non-optimized simple lens (a) which is able to focus down the beam
by a maximum factor of about 1200. Special shaped einzel-lens (b)
with a design developed by Septier \cite{SEPTIER1960} where the
minimal achievable 1$\sigma$-spot size is more than three times
smaller than the one of the non-optimized einzel-lens. Custom made
einzel-lens (c) which is implemented in our setup due to realisable
electrode shapes with similar focusing properties as the special
shaped lens (b). Aside from a maximum focusing factor of
approximately 3000, the simulation predicts for the decel-accel mode
(middle electrode repulsive) optimal results which is important for
further implementation of spherical aberration correction and
utilization of the lens as an ion deflector. Voltages are applied to
the centre electrode (light gray) while the outer ones are grounded
(dark gray).}} 
\label{fig:fig4}
\end{center}
\end{figure}
\emph{Comparison of different lens designs:} 
In order to optimize
the properties of the lens, different shapes are imaginable and have
been discussed
\cite{HAWKES1989,LIEBL2008,SHIMIZU1983,RIDDLE1978,SEPTIER1960}.
Generally, the most important feature is a perfect alignment of the
electrodes in connection with voltage stability which depends, on
the one hand, on the voltage supply, and on the other hand on the
insulating material between the electrodes. For the properties of
the lens it can be presumed that the smaller the dimensions of the
lens, the better its focusing properties are. However, the
decreasing of the lens dimension is limited by the extension of the
ion beam because spherical aberration effects become stronger when
the size of the beam approaches the lens diameter. To assure that
all ions easily go through the lens and spherical aberration is not
dominant, we designed lenses with a main aperture of 1~mm. Usually
the lens' properties improve when the first electrode is thinner
than the second one and the gap spacing between them is minimal. On
the contrary, the distance between the second and the third
electrode has to be much larger to minimize the focal spot size (see
Fig.~\ref{fig:fig4}(a) and (c)). Another possibility to improve the
focusing properties is to change the shape of the electrodes e.g. to
a special design developed by Septier \cite{SEPTIER1960}. The design
is based on a lens with hyperbolic field distribution which has good
imaging characteristics even for steeply inclining beams, but has
been modified to reduce spherical aberration (see
Fig.~\ref{fig:fig4}(b)). Each of the three electrodes has a
different aperture and a completely different shape. For an optimal
adjustment of the lens to our single ion source we have developed a
huge range of different lenses and simulated the achievable spot
sizes.
In Figure \ref{fig:fig4} we present three different types to
demonstrate the enormous influence of the lens design, each with the
focal distance and the appropriate optimized focal spot size at
different voltages and temperatures. A non-optimized einzel-lens
made out of three equivalent electrodes and gaps in between in the
decel-accel mode (see Fig.~\ref{fig:fig4}(a)) allows to focus down
the non deflected ion beam to a 1$\sigma$-spot radius of 25.9~nm
with 2~mK cooled ions (5.1~nm @ 100$\mu$K). Slightly better results
are obtained in the accel-decel mode, where the 1$\sigma$-spot size
produces 20.7~nm for ion beams of 2~mK cooled ions (3.0~nm @
100~$\mu$K). Although this simple einzel-lens already achieves
nanometre spot sizes, the required voltages, especially in the
accel-decel mode, are too high to ensure the isolating property of
the spacer between the electrodes. On the contrary, the lens in
Figure \ref{fig:fig4}(b), which resembles the one developed by
Septier, is able to achieve (in decel-accel mode) a focal
1$\sigma$-spot radius of 6.3~nm, even with ions of 2~mK temperature
before extraction. With ions cooled to the motional ground state
(100~$\mu$K) the simulated spot size decreases down to 1.7~nm. Even
better are the results for the accel-decel mode where the
1$\sigma$-spot size amounts to 5.5~nm for 2~mK and only 0.9~nm for
100~$\mu$K. This proves the desired qualities of our achieved ion
beam and shows explicitly that our setup is able to operate at nm
resolution. However, the realisation of this lens is quite difficult
because of different apertures, the inclined electrodes and the
rounded edges. For this reason we have developed a design with
similar focusing properties but which is easily realisable (see
Fig.~\ref{fig:fig4}(c)).

\begin{figure}[htb]
\begin{center}
\includegraphics[width=14cm]{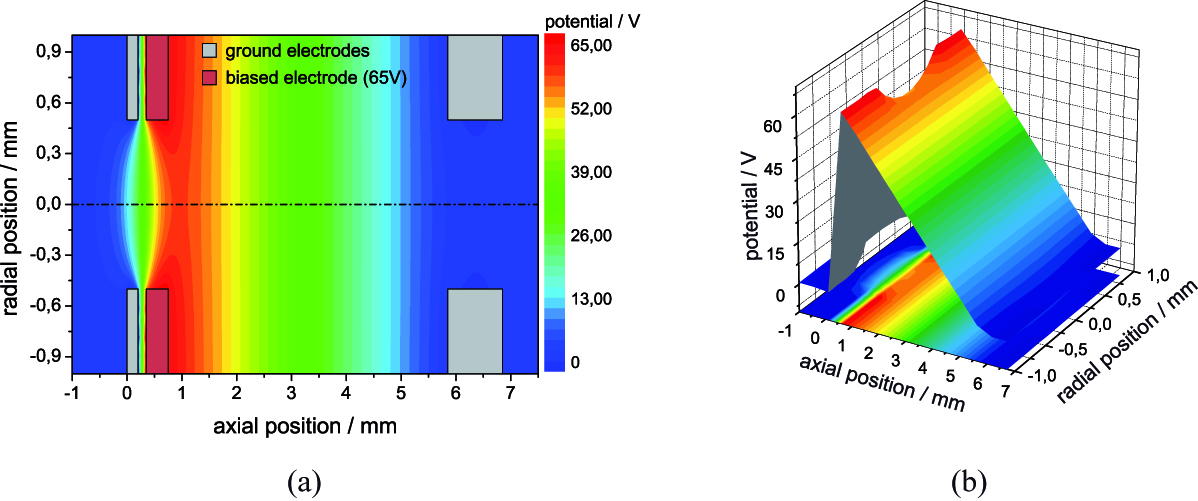}
\caption{\setlength{\baselineskip}{11pt}{\small (a) Potential of the lens from Fig.~\ref{fig:fig4}(c) which
is biased to 65~V. (b) Three dimensional view of the potential
distribution.}} \label{fig:fig5}
\end{center}
\end{figure}
\emph{Analysis of the developed einzel-lens:} The apertures of all
three electrodes of the our developed einzel-lens equal 1~mm, which
facilitates exact alignment. The gaps between the electrodes are
150~$\mu$m wide to ensure the required insulation. For the
accel-decel mode it is possible to reach a 1$\sigma$-spot size of
1.2~nm for 100~$\mu$K cooled ions (1.5~nm @ 2~mK) which is slightly
above the results for the special shaped lens from Septier. Although
the outcomes in the accel-decel mode for 2~mK are better than in the
decel-accel mode, which means this mode shows bigger chromatical
aberration effects, we have chosen to use the lens in the
decel-accel mode to be able to implement spherical aberration
correction and to utilise the lens as an switchable deflector for
ions. With 2~mK cooled ions inside the trap we can achieve a
1$\sigma$-spot size of the extracted ions after focusing through the
lens of 11.0~nm and 1.2~nm for ground state cooled ions,
respectively.

In Figure \ref{fig:fig5}, the induced potential for the lens of
Fig.~\ref{fig:fig4}(c) is plotted when the centre electrode is
biased to 65~V. The potential plot illustrates the function of an
einzel-lens in the decel-accel mode. Before the ions enter the lens,
they are slowed down and deflected slightly away from the axis.
After further defocusing, the ions reach the highest point of the
potential which is approximately 60~V and therefore have the slowest
velocity. Afterwards the ions are accelerated out of the lens and
focused strongly to the axis. The focusing effect relies on the
different ion velocities inside the lens potential. The longitudinal
velocity of the ion is higher during the defocusing period while
climbing to the maximum potential and it is lower during the
focussing period thereafter. This generates the focusing force even
for complete symmetric lens constructions as Fig.~\ref{fig:fig4}(a),
and explains the lower voltages for lens designs where the focusing
space is stretched by positioning the last electrode further away
from the centre electrode (see Fig.~\ref{fig:fig4}(b) and (c)).
\begin{figure}[htb]
\begin{center}
\includegraphics[width=13cm]{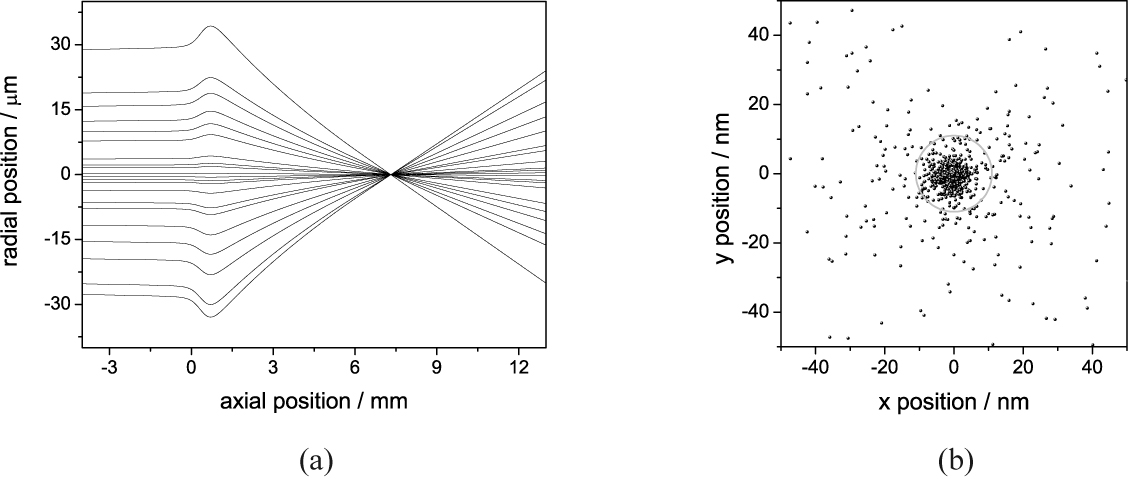}
\caption{\setlength{\baselineskip}{11pt}{\small (a) Trajectories of extracted $^{40}\mathrm{Ca}^+$ ions at
2~mK which are focused by the lens of Fig.~\ref{fig:fig4}(c) in the
decel-accel mode (65~V) to a 1$\sigma$-spot size of 11~nm. (b)
Resulting spot diagram in the focal spot around 0.5~mm after the
last electrode of the lens (light grey circle illustrates the
1$\sigma$-spot radius).}} \label{fig:fig6}
\end{center}
\end{figure}
In Figure \ref{fig:fig6}, trajectories of extracted
$^{40}\mathrm{Ca}^+$ ions through the lens are shown, as well as the
focal spot at a distance of 7.33~mm after entering the lens. When
the ions are extracted with 2~mK initial temperature and the lens is
biased to 65~V, the focal spot occurs half a millimetre after the
last electrode and shows a 1$\sigma$-spot size of only 11~nm.
Reducing the voltage of the centre electrode moves the focal spot
size further behind the lens almost without impairment of the
focusing properties. Lowering the voltage, for instance to 50~V,
generates the spot 8.5~mm after the last electrode, but the spot
size is hardly affected and remains at 12~nm. Therefore, the focal
distance can be adjusted over a few~mm by simply changing the
voltages of the centre electrode. However, the varying range should
be as small as possible since further decreasing of the voltage
increases the spot size (see Fig.~\ref{fig:fig7}(a)). A further
reduction to 25~V, for example, enlarges the spot size up to 52~nm
by a movement of around 100~mm.
\begin{figure}[htb]
\begin{center}
\includegraphics[width=13cm]{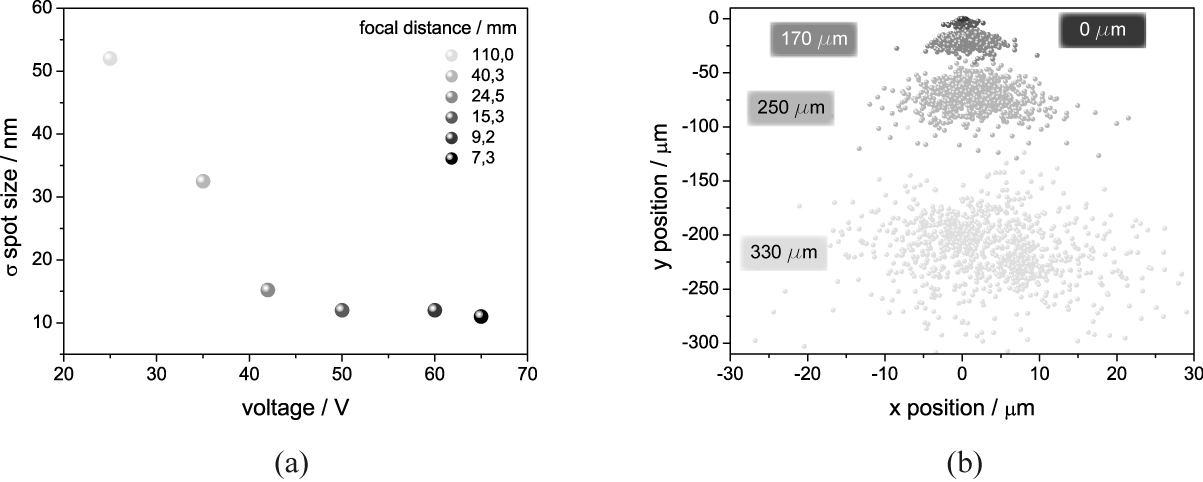}
\caption{\setlength{\baselineskip}{11pt}{\small (a) Enlargement of the 1$\sigma$-spot size by applying
smaller voltages to the centre electrode of the lens. The focal
plane moves away from the lens. 65~V is the upper limit for the lens
voltage because even higher voltages would generate a focus of the
beam inside the lens. (b) If the initial beam is dislocated out of
the centre of the lens, the focusing properties of the lens worsen
or even annihilate. Here, the focal plane is kept constant at 0.5~mm
after the lens as well as the voltage of 65~V. The incoming beam is
displaced upward in the positive x-axis to different distances.}}
\label{fig:fig7}
\end{center}
\end{figure}
As described above, our ion source is able to deflect the ions
during extraction by biasing the deflection electrodes. Small
discrepancies in the alignment of the setup can be corrected by
slight deflection of the beam into the optical axis. The importance
of shooting through the lens without even small initial dislocation
of the ion beam is shown in Figure \ref{fig:fig7}(b). For example, a
displacement of the incoming beam of half the aperture radius
(250~$\mu$m) annihilates the focusing property. Besides the
displacement of the outgoing beam in the opposite direction, it also
diverges to similar spot dimensions as it has without any ion
optics. With further shifting of the ion beam the lens even enlarges
the spot size in the intended focal plane. Hence, for this required
precise alignment, the deflection electrodes present the best way to
aim the extracted ion beam to the centre of the lens aperture. A
displacement in one direction of the trap axis of 250~$\mu$m for
example is neutralized by applying a potential of 0.36~V to two
opposing trap blades and leads only to a small increase of the
1$\sigma$-spot size from 16.1~$\mu$m to 17.2~$\mu$m. However, when
our setup is adjusted accurately, it is, in combination with the
specially developed einzel-lens, a perfect implantation source,
where nearly every species of charged particles or molecules can be
deterministically implanted to any substrate with nanometre
resolution.

\begin{figure}[htb]
\begin{center}
\includegraphics[width=15cm]{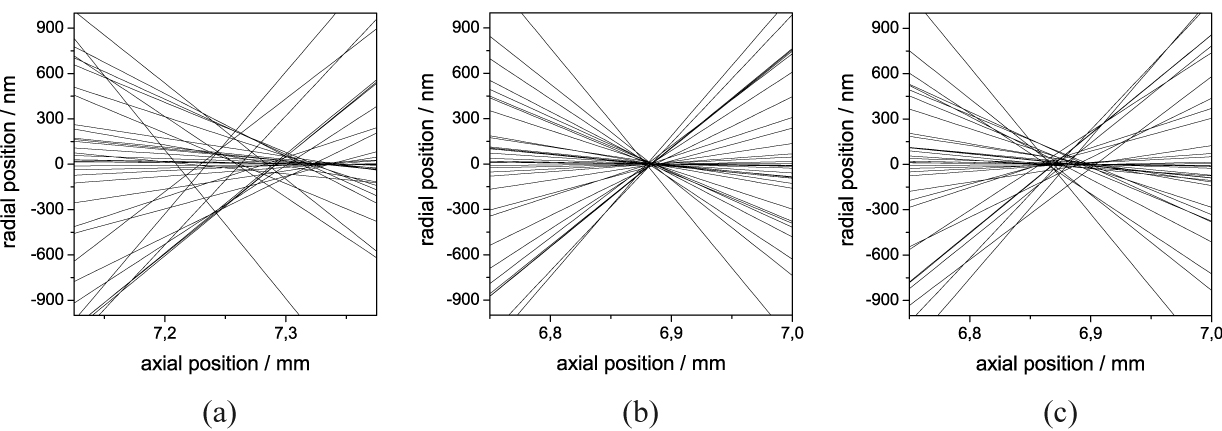}
\caption{\setlength{\baselineskip}{11pt}{\small Zoom of the focal regions to show the effect of spherical
aberration correction. All diagrams were calculated with an incoming
beam with a 1$\sigma$-spot radius of 36~$\mu$m. For the first two
diagrams (a) and (b) the chromatical aberration effect is
artificially removed by setting the velocity fluctuations of the
ions to zero. The uncorrected spot (a) is achieved with the lens
constantly biased to 65~V and shows a 1$\sigma$-spot radius of 52
nm. On the contrary, the spherical corrected spot (b) only has
1$\sigma$-size of 12~nm. 210~ns after the ions has entered the lens,
the voltage of the centre electrode is switched from 35~V to 85~V.
The same incoming beam and correction settings of the lens are used
for the trajectories in diagram (c), only with an added velocity
fluctuation of 6.3~m/s to show the negative effect of the chromatic
aberration of the lens. Here, 68 \% of all simulated ion
trajectories lie within a spot radius of 57~nm.}} \label{fig:fig8}
\end{center}
\end{figure}
\emph{Correction of the spherical aberration:} One of the most
advantageous properties of our setup is the fact that the ion
extractions are synchronised to the rf voltage and the relative
velocity uncertainty is extremely low. Thus, the ions are well
defined in space and time during the whole extraction, as well as
during the focusing process. With that, it is possible to switch the
lens to different voltages at different points in time, which
enables the correction of aberration effects.
When compared to the lens from Septier our design shows a larger
spherical aberration which enlarges the spot size a little bit.
According to Scherzer's theorem, it is not possible to avoid such
spherical and chromatical aberration with rotationally symmetric
electrostatic einzel-lenses in a charge free space
\cite{SCHERZER1936}. Another way of interpreting the theorem is by
stating the impossibility of the realisation of a diverging lens.
One way of circumventing this theorem is to use time dependent
electrostatic fields \cite{SCHOENHENSE2002} simply by switching the
lens to another voltage at a well defined time. This leads to forces
a diverging lens would create and therefore spherical aberration can
be reduced. Because of a fast increase of the lens potential at a
well defined time, the particle is accelerated and therefore
slightly diverged. Ions with outer trajectories react stronger on
the potential shift and subsequently intersect the optical axis
further away. The switching time, where the lens voltage is altered
from one value to another, has to be accurately chosen and has to be
as short as possible. With our high voltage switches mentioned above
we are able to vary the voltage within 5~ns, which should be short
enough to neglect additional energy broadening due to finite rise
time of the electric field. However, the spherical aberration
correction has to be adjusted to the incoming single ion beam.
Simulation shows that for 2~mK cooled ions (which means a
1$\sigma$-spot size of 16.1~$\mu$m) the best results are obtained
when the lens is primarily grounded then biased to 60~V, 170~ns
after the ions have passed the first aperture . Thus the
1$\sigma$-spot in the focal plane can be nearly reduced by a factor
of 2 from 11~nm down to 6.0~nm. The same switching time and voltage
reduces the focal 1$\sigma$-spot size for ground state cooled ions
from 1.2~nm to 0.9~nm. If the ion beam dimensions are increased, the
factor of improvement also increases slightly. For example, beam
characteristics similar to the ones we expect in our experiment
after perfectly aligning the lens axis to the trap axis
(1$\sigma$-spot of 36~$\mu$m and a maximum velocity uncertainty of
only 6~m/s) requires a different setting for the switching time and
voltage, but the reduction of the focal spot increases by more than
a factor of 2 (see Fig.~\ref{fig:fig8}). Here, the voltage is
switched from 35~V during the entering of the ions to 85~V after
210~ns. The focused spot amounts to 138~nm when the lens is
constantly biased to 65~V and decreased down to 57~nm for the
spherical aberration corrected lens due to switching of the voltage
(see Fig.~\ref{fig:fig8}(c)). The remaining spread of the focal spot
is therefore mainly due to the remaining chromatic aberration of the
lens because of the velocity fluctuation. In order to show solely
the enhancement of the correction of the spherical aberration, the
velocity fluctuations have been artificially removed in Figure
\ref{fig:fig8}(a) and \ref{fig:fig8}(b). Here, the spot improves
with the spherical aberration correction from a 1$\sigma$-spot size
of 52~nm down to 12~nm. Note that the spot in Figure
\ref{fig:fig8}(a) does not show an accurate Gaussian distribution
and therefore the measured value of 68\% within 52~nm cannot
obviously be read out from the plot. But Figures \ref{fig:fig8}(a)
and \ref{fig:fig8}(b) give a good insight of the enhancement due to
spherical aberration correction, whereas Figure \ref{fig:fig8}(c)
shows the effect of chromatic aberration. So far chromatic errors
will not be corrected by the presented ion optics, but this can be
implemented with time dependent electric fields
\cite{SCHOENHENSE2002}.

\begin{figure}[htb]
\begin{center}
\includegraphics[width=7.5cm]{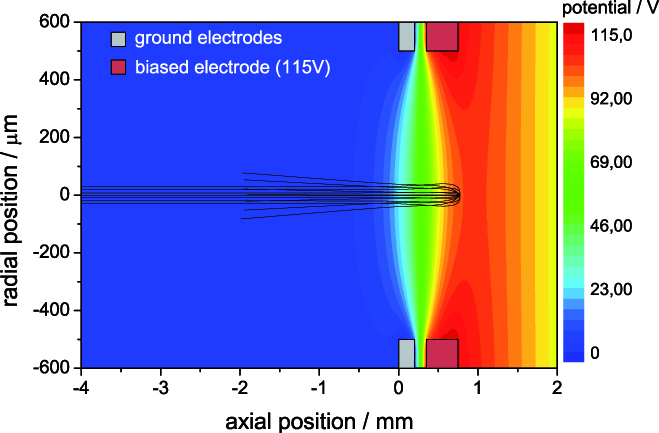}
\caption{\setlength{\baselineskip}{11pt}{\small Trajectories of reflected $^{40}\mathrm{Ca}^+$ ions and
potential distribution of the lens set to 115~V at the centre
electrode (red).}}
\label{fig:fig9}
\end{center}
\end{figure}
\emph{Reflection of the cooling ions and possible post
acceleration:} Another advantage of the triggered extraction is the
possibility to reflect the cooling ions during the implantation
process. Because of unavoidable heating processes during a possible
separation of the ions inside the trap \cite{ROWE2002}, it is
preferable to separate the cooling $^{40}\mathrm{Ca}^+$ ions from
the implantation particles by reflecting the cooling ions at the
einzel-lens. Due to different masses of the additionally loaded,
sympathetically cooled ions compared to the $^{40}\mathrm{Ca}^+$
ions, the flight velocities differ from each other. Therefore, the
cooling ions arrive at different times at the lens and can be easily
reflected by applying higher voltages. Nitrogen, for instance, is
accelerated in the simulation up to 36.2~km/s due to its smaller
mass, and thus arrives approximately 4.3~$\mu$s earlier at the lens
than the $^{40}\mathrm{Ca}^+$ ions.
Hence, the lens has to be switched to a higher potential after the
nitrogen particles have passed. The required voltage for reflection
has to be at least 115~V, which is experimentally feasible (see
Fig.~\ref{fig:fig9}). Also worth mentioning is that
post-acceleration will even improve the focusing results by reducing
the chromatical aberration due to a smaller relative velocity
fluctuation. A simple cylindrical tube positioned in the ion beam
axis and biased to 10 kV shows promising focusing effects when
switched off after the ions are inside the tube. First simulated
results predict velocities above 200~km/s and focal 1$\sigma$-spot
sizes of less than 4 $\AA$ for ground-state cooled ions. However,
switching times have to be even more precise due to higher
velocities and therefore shortened timescales. Moreover, the gained
reduction of the spatial resolution is partly neutralised by
straggling effects of the implanted ions inside the target material.

\section{Experimental results of the novel ion source}
\begin{figure}[htb]
\begin{center}
\includegraphics[width=14cm]{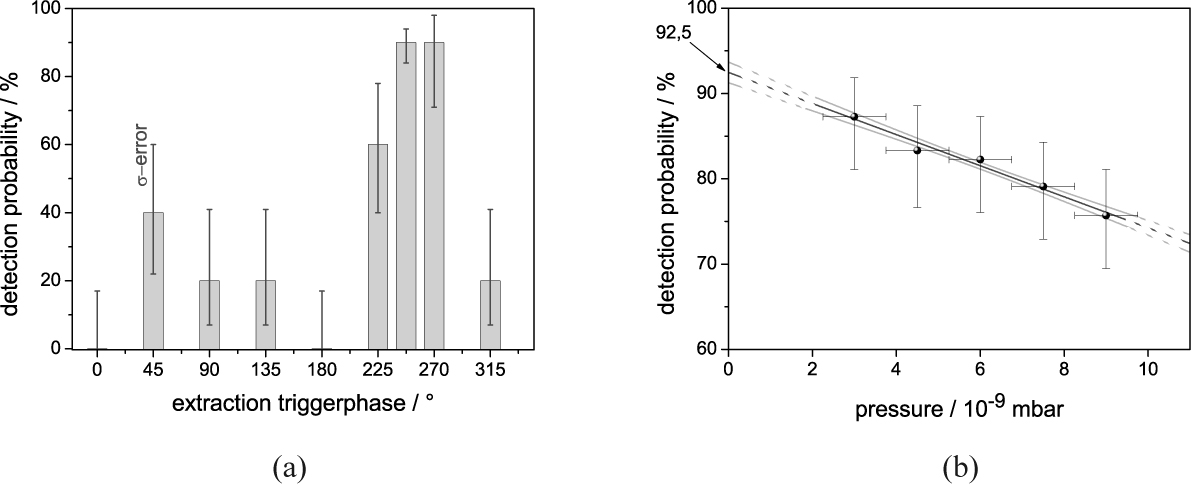}
\caption{\setlength{\baselineskip}{11pt}{\small (a) Dependence of detection probability on different
extraction trigger phases. (b) Detection probability during the
extraction of single ions while the pressure changes. A linear fit
is added (black line) as well as the standard deviation (grey line).
The measurement is based on 251 successful extractions out of 310
shots. Both diagrams are measured without any aperture in front of
the detector.}} \label{fig:fig10}
\end{center}
\end{figure}
We have implemented a linear segmented Paul trap as a novel ion
source which is deterministic and has promising characteristics in
velocity fluctuations and spot sizes for focusing the beam to
nanometre resolution \cite{SCHNITZLER2009}. As mentioned above, the
source has to be accurately synchronised to the phase of the rf trap
drive voltage. We have therefore developed electronics that
synchronise the TTL signal from the lab computer to the rf phase.
The dependence of the hit rate on the extraction trigger phase is
shown in Figure \ref{fig:fig10}(a) where the detection probability
has a maximum at a rf phase of around 250$^\circ$ to 270$^\circ$.
During all these measurements, the deflection voltages were kept
constant and no additional apertures were placed in front of the
detector (entrance aperture of the detector equals 20~mm). With
smaller apertures, the contrast between detecting 90 \% of the
extracted ions and missing the detector would be even stronger.
Additionally, the experiment shows a weak dependence on the pressure
in the vacuum chamber (see Fig.~\ref{fig:fig10}(b)). When improving
the vacuum down to a few 10$^{-9}$~mbar, the detection rate again
improves to almost 90 \% as well.  A linear extrapolation leads to a
maximum detection probability of 92.5 \%. The dependency on the
pressure is mainly based on possible collisions with the background
residual gas particles. Furthermore, the stability of the ions in
the trapping potential is lessened with higher pressure, which leads
to higher fluctuations of the extraction trajectories. However, our
measured hit-rate lies above the specified quantum efficiency of the
detector (80 \%).

\begin{figure}[htb]
\begin{center}
\includegraphics[width=14cm]{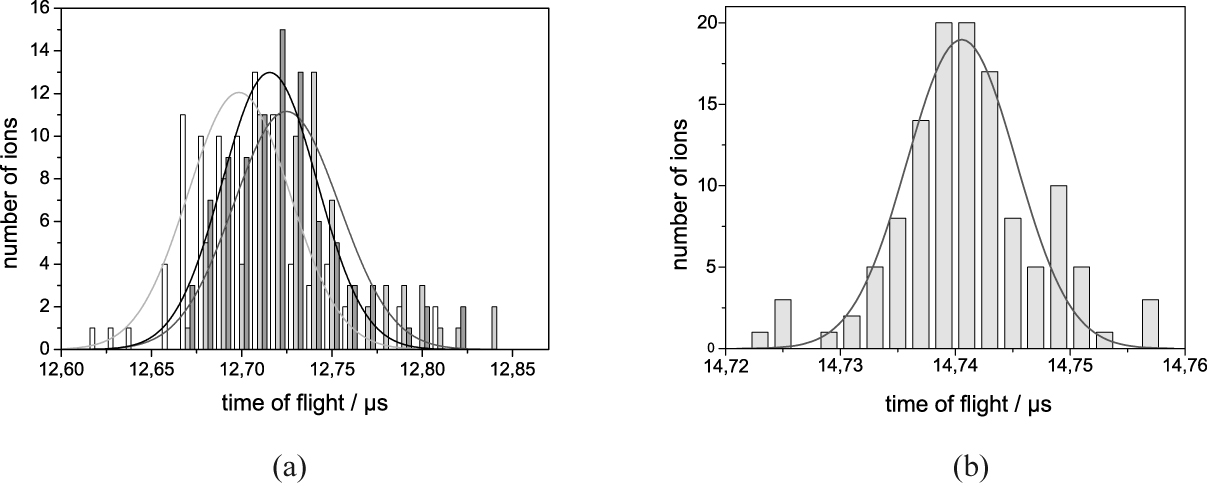}
\caption{\setlength{\baselineskip}{11pt}{\small (a) Time of flight spectrum for two-ion crystals based on
93 successful detections without any aperture plates in front of the
detector.The mean value for the first detected ion (white bins)
amounts to 12.699~$\mu$s with a 1$\sigma$-spread of 56~ns (light
grey line) and for the second ion (grey bins) to 12.725~$\mu$s with
a 1$\sigma$-spread of 59~ns respectively (grey line). The difference
in arrival time between the first and the second ion amounts to
$\Delta \bar t$= 26~ns. Additionally, the mean value (dark grey
bins) for the time of flight for both detection events adds up to
12.715 $\mu$s with a 1$\sigma$-spread of 52~ns (dark grey line). The
bin size of the histogram is 10~ns. (b) Time of flight spectrum for
single ion crystals based on 123 successful detections out of 139
shots through the 1~mm aperture. Gaussian fit of the data leads to
an average time of 14.74~$\mu$s with a 1$\sigma$-spread of 9.5~ns.
The bin size of the histogram equals 2~ns. Note that the different
average times in (a) and (b) are caused by a repositioning of the
detector backwards due to the installation of an aperture plate, and
that the improvement of the 1$\sigma$-spread is mainly due to the
enhancement of the electronics of the phase delay trigger.}}
\label{fig:fig11}
\end{center}
\end{figure}
\emph{Determination of the velocity fluctuation:} In order to get
information concerning the velocity fluctuation of the extracted
ions we have measured the time of flight of each detected ion or ion
crystal. The time of flight spectrum for ion crystals consisting of
two ions is depicted in Figure \ref{fig:fig11}(a). The difference
between the arrival times of the first and the second ion results in
$\Delta \bar t$= 26.3~ns, which is mainly based on the Coulomb
interaction. The spectrum shows that the ions remain in their
crystalline structure during the extraction, which is important for
the above mentioned post separation of the $^{40}\mathrm{Ca}^+$ ions
from possible dopant ions or the implantation of a whole crystal
with one extraction event. In order to characterize the spatial
divergence of the ion beam, a movable aperture plate was installed
in front of the detector. Along with this modification, a faster and
more stable trigger phase electronic unit was also implemented. The
subsequent measurements of the time of flight spectra are shown in
Figure \ref{fig:fig11} (b). With the improvement of the phase delay
trigger the 1$\sigma$-width reduces from the previous 50~ns down to
9.5~ns. The movement of the detector further away from the trap of
about 30~mm causes an elongation of the time of flight from around
12.7~$\mu$s up to 14.74~$\mu$s. Therefore, the mean velocity of the
extracted single $^{40}\mathrm{Ca}^+$ ions is measured to be
19.47~km/s with a 1$\sigma$-width of 6.3~(6)~m/s, meaning that the
velocity uncertainty is a factor of 10 larger than inside the trap
for the 2~mK cooled ions. The relative velocity uncertainty
$\Delta$v/v is only 3.2$\times $10$^{-4}$ with the possibility of
further reduction due to post-acceleration.

\begin{figure}[htb]
\begin{center}
\includegraphics[width=15.5cm]{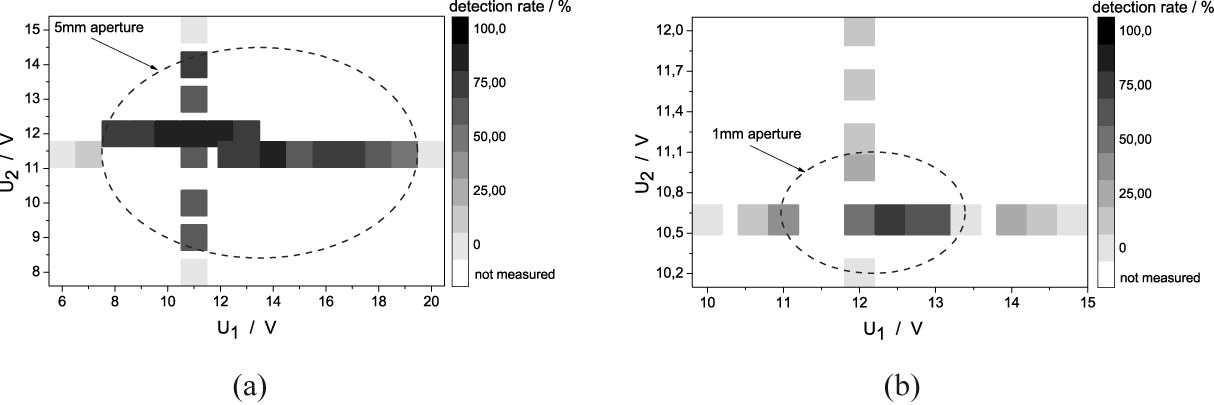}
\caption{\setlength{\baselineskip}{11pt}{\small Detection probabilities as a function of the deflection
voltages. U$_1$ and U$_2$ indicate the potential applied between two
opposing blades. The scan clearly shows the edge of the aperture by
a sharp decrease of the detection rate. (a) The range can be varied
over 10~V and 5~V for the 5~mm aperture. (b) For the 1~mm hole the
range where ions can be efficiently detected decreases to 2~V and
1~V respectively.}} \label{fig:fig12}
\end{center}
\end{figure}
\emph{Determination of the spatial beam dimension:} The beam
divergence is determined by scanning the beam over different
apertures of 5~mm, 1~mm and 300~$\mu$m diameter by altering the
deflection voltages. Measurements conducted with the 5~mm aperture
have shown that over a voltage range of 5~V between two opposing
blades and 10~V between the other two opposing blades we are even
able to detect ion crystals of up to 6 ions with an efficiency above
90\% (see Fig.~\ref{fig:fig12}(a)). The edge of the aperture is also
clearly recognizable in the scan diagram due to a sharp decrease of
the detection rate. When reducing the aperture diameter to 1~mm the
scanning range of the deflection voltages are scaled down to 1~V and
2~V respectively (see Fig.~\ref{fig:fig12}(b)). In the centre of the
scanned voltage range we detect 87\% of the single extracted ions
while the detection of ions which are extracted in a crystal shows a
reduced detection rate of 78\%. Note that the detection rate of
nearly 90\% for single ions lies above the specified quantum
efficiency of the the detector and is mainly limited by its
performance. However, further downsizing of the dimension of the
aperture to 300~$\mu$m reduces the detection rate to 68\% which
indicates the cut off of the spot by the edge of the aperture.
\begin{figure}[h]
\begin{center}
\includegraphics[width=14cm]{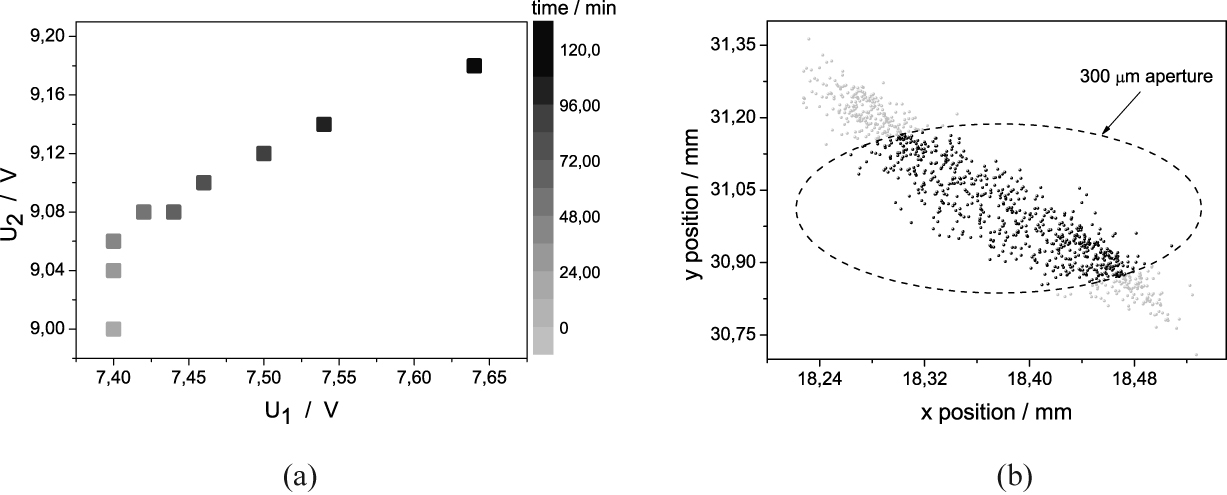}
\caption{\setlength{\baselineskip}{11pt}{\small (a) Variation of optimal voltages U$_1$ and U$_2$ for the
two pairs of opposing blades during measurements of two hours
duration. The change in voltages compensates for the drift of the
ion beam. Therefore, the detection probability could be kept
constant at 68\%. (b) Simulation of the spot with experimental
settings. Black spots illustrate ions that would have passed the
300~$\mu$m aperture whereas grey spots are blocked by the aperture.
Therefore, 65\% of the extracted ions can be detected after the
aperture which is in good agreement with our experimental result (68
\%). Note that the spot is not a Gaussian distribution and is
displaced by a few millimetres due to deflection voltages U$_1$ =
7.5~V and U$_2$ = 9.1~V.}}
\label{fig:fig13}
\end{center}
\end{figure}
By scanning over the 5~mm aperture at two different times we
observed a drift of the ion beam of 15~$\mu$m/min, possibly caused
by temperature drifts or a displacement of the ions due to patch
electric fields on the insulating surface between the electrodes.
When using the 300~$\mu$m aperture, the drift becomes more
significant, which we compensated by adjusting the deflection
voltages (see Fig.~\ref{fig:fig13}).

\emph{Comparison of the experiment and the simulation:} For a
comparison of the experimental results with the findings from the
simulations it is useful to obtain the simulated spot shape with all
experimentally used settings. The resulting simulated spot again
shows an oval shape. The velocities have a Gaussian distribution
(similar to the experiment) with a 1$\sigma$-spread of only 2.7~m/s.
This means the experimentally measured values are by a factor of 2.2
worse than the simulated results. If it is assumed that the same
holds for the spatial spreading of the spot, it is possible to
compare the results of the experiment and the simulation. With the
cigar shaped spot that is enlarged by a factor of 2.2, it would be
possible to get 65\% of the extracted ions through the 300~$\mu$m
aperture (see Fig.~\ref{fig:fig13}(b)). In our experiment we are
able to detect 68\% of the single ions after the 300~$\mu$m hole,
with adjustment of the deflection voltages during the extractions.
From the simulation, we can also deduce that our trap is tilted by
an angle of 4.2$^\circ$ at the x-axis and 7.2$^\circ$ for the
y-axis.

\section{Outlook}
We have shown that our setup is able to deterministically extract
single ions on demand. With additional optimized ion optics, the
numerical simulation predicts a spot size with nm dimension. For an
experimental confirmation of the predicted focal spot, different
methods are conceivable. Similar to the previous measurements of the
spatial beam divergence, it is possible to implement an aperture
with less than 30~nm, which can be drilled, for example, by a
focused ion beam (FIB) \cite{MEIJER2006,MEIJER2008}. Another
possibility to confirm the expected nm resolution of our combined
system of ion trap and ion optics is to measure the achieved
resolution after implanting the ions into solid state surfaces. With
a STM, it is possible to identify subsurface impurities with nm
resolution, produced by cobalt particles in a copper bulk material,
for example \cite{WEISMANN2009}. A further method to confirm the
resolution of single implanted ions makes use of generated NV colour
centres in diamond \cite{NEUMANN2008}. Dipolar coupling of single
nuclei localized close to the defect can be used to measure the
achieved implantation resolution.

\section*{Acknowledgment}
We acknowledge financial support by the Landes\-stiftung
Baden-W\"urttemberg in the framework 'atomics' (Contract No. PN
63.14) and the 'Eliteprogramm Postdoktorandinnen und
Postdoktoranden', the European commission within EMALI (Contract No.
MRTN-CT-2006-035369) and the VolkswagenStiftung.

{}

\end{document}